\documentclass[amsmath,natbib,rotating]{mn2e}
\usepackage{graphicx}
\title[Gas phase formation of formamide]{Gas phase formation of the prebiotic molecule formamide: insights from new quantum computations}

\author[V. Barone et al.]{
V. Barone,$^{1}$\thanks{E-mail: vincenzo.barone@sns.it}
C. Latouche,$^{1}$\thanks{E-mail: camille.latouche@sns.it}
D. Skouteris,$^{1}$\thanks{E-mail: dimitrios.skouteris@sns.it} 
F. Vazart,$^{1}$\thanks{E-mail: fanny.vazart@sns.it}
N. Balucani$^{2,3,4}$\thanks{E-mail: nadia.balucani@unipg.it}
\newauthor
C. Ceccarelli$^{3,4}$\thanks{E-mail: Cecilia.Ceccarelli@obs.ujf-grenoble.fr} 
B. Lefloch$^{3,4}$\thanks{E-mail: Bertrand.Lefloch@obs.ujf-grenoble.fr} 
\\
$^{1}$Scuola Normale Superiore, Piazza dei Cavalieri 7, 56126 Pisa, Italy\\
$^{2}$Dipartimento di Chimica, Biologia e Biotecnologie, Via Elce di Sotto 8, 06123 Perugia, Italy\\
$^{3}$ Univ. Grenoble Alpes, IPAG, F-38000 Grenoble, France \\
$^{4}$ CNRS, IPAG, F-38000 Grenoble, France\\
}

\begin{document}

    \date{Received - ; accepted -}

\pagerange{\pageref{firstpage}--\pageref{lastpage}} \pubyear{2015}

\maketitle

\label{firstpage}

\begin{abstract} New insights into the formation of interstellar
  formamide, a species of great relevance in prebiotic chemistry, are
  provided by electronic structure and kinetic calculations for the
  reaction NH$_2$ + H$_2$CO $\rightarrow$ NH$_2$CHO + H. Contrarily to
  what previously suggested, this reaction is
  essentially
  barrierless and can, therefore, occur under the low temperature
  conditions of intestellar objects thus providing a facile formation
  route of formamide. The rate coefficient parameters for the reaction
  channel leading to NH$_2$CHO + H have been calculated to be
  $A = 2.6 \times 10^{-12}$ cm$^3$ s$^{-1}$, $\beta = -2.1$ and
  $\gamma = 26.9$ K in the range of temperatures 10-300 K. Including
  these new kinetic data in a refined astrochemical model, we show
  that the proposed mechanism can well reproduce the
  abundances of formamide observed in two very different interstellar
  objects: the cold envelope of the Sun-like protostar IRAS16293-2422
  and the molecular shock L1157-B2. Therefore, the major conclusion of
  this Letter is that there is no need to invoke grain-surface
  chemistry to explain the presence of formamide provided that its
  precursors, NH$_2$ and H$_2$CO, are available in the gas-phase.
\end{abstract}


%%%%%%%%%%%%%%%%%%%%%%%%%%%%%%%%%%%%%%%%%%%%%%%%%%

%%%%%%%%%%%%%%%%% BODY OF PAPER %%%%%%%%%%%%%%%%%%

\section{Introduction}

Interstellar molecules including a peptide bond -NH-C(=O)-, such as
formamide (NH$_2$CHO) and acetamide, (NH$_2$COCH$_3$), are
particularly interesting for their potential role in prebiotic
chemistry (Saladino et al. 2012). Formamide was detected for the first
time in space towards Sgr B2 by Rubin et al. (1971) and later in Orion
KL (Turner 1989). Since then, it has been observed in several massive
hot cores (Bisschop et al. 2007; Adande et al. 2011), low-mass hot
corinos (Lopez-Sepulcre et al. (2015), the cold envelope and hot
corino of IRAS16293-2422 (Kahane et al. 2013), and the outflow shock
regions L1157-B1 and B2 (Yamaguchi et al. 2012; Mendoza et al. 2014).
Formamide was also detected in the Hale-Bopp comet (Bockelee-Morvan et
al. 2000) and, more recently, in comets C/2012 F6 and C/2013 R1 (Biver
et al. 2014). Therefore, formamide is present in a large variety of
star-forming environments, as well as in Solar System comets, thus
supporting the hypothesis that molecules with a strong prebiotic
potential could have been delivered to Earth by comets after being
synthesized in prestellar environments (e.g. Caselli \& Ceccarelli,
2013).

In a very recent study by Lopez-Sepulcre et al. (2015), formamide has
been searched for in ten low- and intermediate-mass pre-stellar and
protostellar objects as a part of the IRAM Large Programme ASAI
(Astrochemical Surveys At IRAM), which makes use of unbiased
broad-band spectral surveys at millimetre wavelengths. While the
related species HNCO (isocyanic acid) has been detected in all
objects, formamide has not been identified in five of them, which are
the coldest and devoid of hot corinos. According to those results,
Lopez-Sepulcre et al. (2015) suggested that HNCO is formed in the gas
phase during the cold stages of star formation, while NH$_2$CHO is
formed from the hydrogenation of HNCO on the ice mantles of dust
grains and remains frozen until the temperature rises enough to cause
the icy grain mantles to sublimate. Nevertheless, very recent
experimental work on the hydrogenation of the frozen HNCO seems to
dispute the suggestion that HNCO is the parent species of NH$_2$CHO on
ice (Noble et al. 2015). Other heterogeneous processes on the icy
surface of interstellar grains have also been considered (Jones et al.
2011; Walsh et al. 2012; Garrod et al. 2008).

The formation routes of NH$_2$CHO in the gas phase have been only
partially investigated. Quan \& Herbst (2007) suggested that the
radiative association reaction between formaldehyde and protonated
ammonia followed by dissociative electron recombination could be a
source of formamide. However, the model of Quan \& Herbst could only
produce an abundance of formamide of $\sim 10^{-15}$ (with respect to
H$_2$), significantly lower than the value found by Hollis et al.
(2006) in Sgr B2. Instead of radiative association, Halfen et al.
(2011) suggested that the ion-molecule reaction between formaldehyde
and protonated ammonia could lead to protonated formamide, which, in
turn, forms formamide by dissociative recombination according to
NH$_3$CHO$^+$ + e$^-$ $\to$ NH$_2$CHO + H.
A recent theoretical study of the reaction NH$_4$$^+$
+ H$_2$CO confuted this suggestion (Redondo et al. 2014a) because it was
verified that all the channels of the potential energy surface leading
to protonated formamide exhibit high-energy barriers. The feasibility
of other gas-phase ion-molecule reactions that could produce
precursors of formamide has been recently explored by Redondo et al.
(2014b) who considered the ion-molecule reactions between
NH$_3$OH$^+$ and NH$_2$OH$^+$ with H$_2$CO
and HCOOH. Also in these cases, the presence of high-energy barriers
along the reaction pathways has been exhibited. 

As for neutral-neutral
reactions, the NH$_2$ + H$_2$CO $\to$ NH$_2$CHO + H seems to be a 
viable route and, indeed, this reaction has been
initially considered in the OSU database (Harada \& Herbst 2008; {\it
  http://faculty.virginia.edu/ericherb}) with an estimated rate
coefficient in the gas kinetics range ($10^{-10}$ cm$^3$ s$^{-1}$),
with the assumption that it is a barrier-less reaction. Later on,
however, Garrod (2013) disregarded such an assumption with the
following reasons. According to the theoretical study by Li \& Lu
(2002), the more exothermic product channel leading to NH$_3$
+ HCO is characterized by an entrance barrier of $5.89$ kcal mol$^{-1}$
($\sim$ 3000 K) with an estimated rate coefficient of $5.25
\times 10^{-17}$ cm$^3$ s$^{-1}$.
Li \& Lu (2002) investigated only the channel leading to NH$_3$
+ HCO. Nonetheless, Garrod (2013), by drawing an analogy with the
reaction OH + H$_2$CO
where the channel leading to HCOOH + H has a branching ratio of 
$\sim 2\%$ compared to the dominant H$_2$O + HCO channel, concluded
that the role of the NH$_2$ + H$_2$CO $\to$ NH$_2$CHO + H must be
irrelevant. The reaction was therefore excluded from the gas-phase
network and rather considered a possible formation route of formamide
when in ice-assisted chemistry (Garrod 2013; Garrod et al. 2008).

In this work, we present state of the art quantum mechanical
characterizations of the characteristic stationary points (minima and
first order saddle points, also referred to as transition states) on
the potential energy surface (PES) governing the NH$_2$ + H$_2$CO
reaction channels (\S 2 and 3). At variance from previous suggestions by
Garrod (2013) and Garrod et al. (2008), the reaction pathway
proceeding with the addition of the nitrogen atom of the NH$_2$ group to
the carbon atom of formaldehyde has been found to be 
essentially barrierless. Actually, the entrance channel is characterized by 
a minimum corresponding to a Van der Waals complex followed by a first order 
saddle point, but both stationary points lie below the asymptotic reactants 
energy according to CBSQB3 calculations. 
We have chosen to treat the Van der Waals region adiabatically rather than 
assume that energy randomization takes place at the minimum, as such a process 
is not expected to be efficient. 
 For
this reason, we have also employed capture theory as well as
Rice-Ramsperger-Kassel-Marcus (RRKM) calculations to derive the
reaction rate coefficient as a function of the temperature in the
range between 10 and 300 K, that is, the range of relevance for the
interstellar objects of interest. The approach is the same recently
used by us to investigate a bimolecular formation route of
cyanomethanimine, another complex organic molecule with a prebiotic
potential (Vazart et al. 2015). The derived rate coefficient has also
been tested in astrochemical models and compared with the formamide
abundance measured in star formation regions (\S 4). A final section
(\S 5) discusses the implications of the computations presented here
and the future perspectives.

%%%%%%%%%%%%%%%%%%%%%%%%%%%%%%%%%%%%%%%%%%%%%%%%%%%%%
\section{Computational details}

\subsection{Electronic calculations}
\label{sec:electronic} % used for referring to this section from elsewhere

All calculations have been performed with a development version of the
Gaussian suite of programs (Frisch et al. 2013). Most of the
computations were performed with the double-hybrid B2PLYP functional
(Grimme 2006) in conjunction with the m-aug-cc-pVTZ basis set
(Papajak et al., 2009; Dunning 1989) where $d$ functions on
hydrogens have been removed. Semiempirical dispersion contributions
were also included by means of the D3BJ model of Grimme (Goerigk \&
Grimme 2011; Grimme et al. 2011). Full geometry optimizations
have been performed for all molecules checking the nature of the
obtained structures (minima or first order saddle points) by
diagonalizing their Hessians. More accurate electronic energies were
obtained by the Complete Basis Set (CBS-QB3) method, which employs a
coupled cluster ansatz including single and double excitations
together with a perturbative estimation of triple excitations
(CCSD(T)) in conjunction with complete basis set extrapolation
(Montgomery et al. 2000; Ochterski et al. 1996). 

\subsection{Kinetic calculations}
\label{sec:kinetic} % used for referring to this section from elsewhere

We have performed kinetic calculations using an in house code
described in previous papers (Leonori et al. 2009, 2013; Vazart et
al. 2015). The initial bimolecular rate constant leading from the
reactants to the intermediate is evaluated using capture theory
calculations, after fitting the energy values for the approaching
reactants to a $1/R^6$  functional form and assuming that each
successful capture leads to the intermediate. 
As far as dissociation back to reactants is concerned, we have used a
detailed balance argument, whereby the unimolecular rate constant for
back-dissociation is given by the equation  
\begin{equation}
  k_{back}= k_{capt}  \frac{\rho(R)}{\rho(I)}
\end{equation}
where $k_{capt}$ is the capture rate constant, $\rho(R)$ is the density of states per unit volume for the reactants and $\rho(I)$ is the density of states for the intermediate. 
For the intermediate dissociation into formamide + H, where, as
opposed to the reactants, a well-defined transition state exists, we
have performed a RRKM calculation. The microcanonical rate constant is
calculated using the formula 
\begin{equation}
  k(E)= \frac{N(E)}{h \rho(E)}
\end{equation}
where $N(E)$ denotes the sum of states in the transition state at
energy $E$, $\rho (E)$ is the reactant density of states at energy $E$
and $h$ is Planck's constant. $N(E)$ is obtained by integrating the
relevant density of states up to energy $E$ and the rigid
rotor/harmonic oscillator model is assumed. Both densities of states
(reactant and transition state) are appropriately symmetrized with
respect to the number of identical configurations of the reactants
and/or transition state.   
Tunneling and quantum reflection have been taken into account by
computing the tunneling probability for an Eckart barrier having the
same negative second derivative at the maximum of the pertinent saddle
point. 
After all calculations have been performed, the branching ratio
between products and back-dissociation is determined for each energy
and the corresponding capture rate constant is multiplied by this
ratio to give the rate constant for the formation of formamide.
Finally, the rate constants are Boltzmann-averaged in order to provide
the rate constants as a function of temperature.

\section{Results}

\subsection{Electronic and vibrational investigations} 

The structures of all the molecules were optimized at the
B2PLYP-D3/m-aug-cc-pVTZ and CBS-QB3 levels of theory. All the
precursors, intermediate and products were fully characterized as
minima on the potential energy surface and transition states exhibited
a single imaginary frequency thanks to vibrational calculations at
B2PLYP-D3/m-aug-cc-pVTZ level.
\begin{figure} \includegraphics[width=0.95\columnwidth]{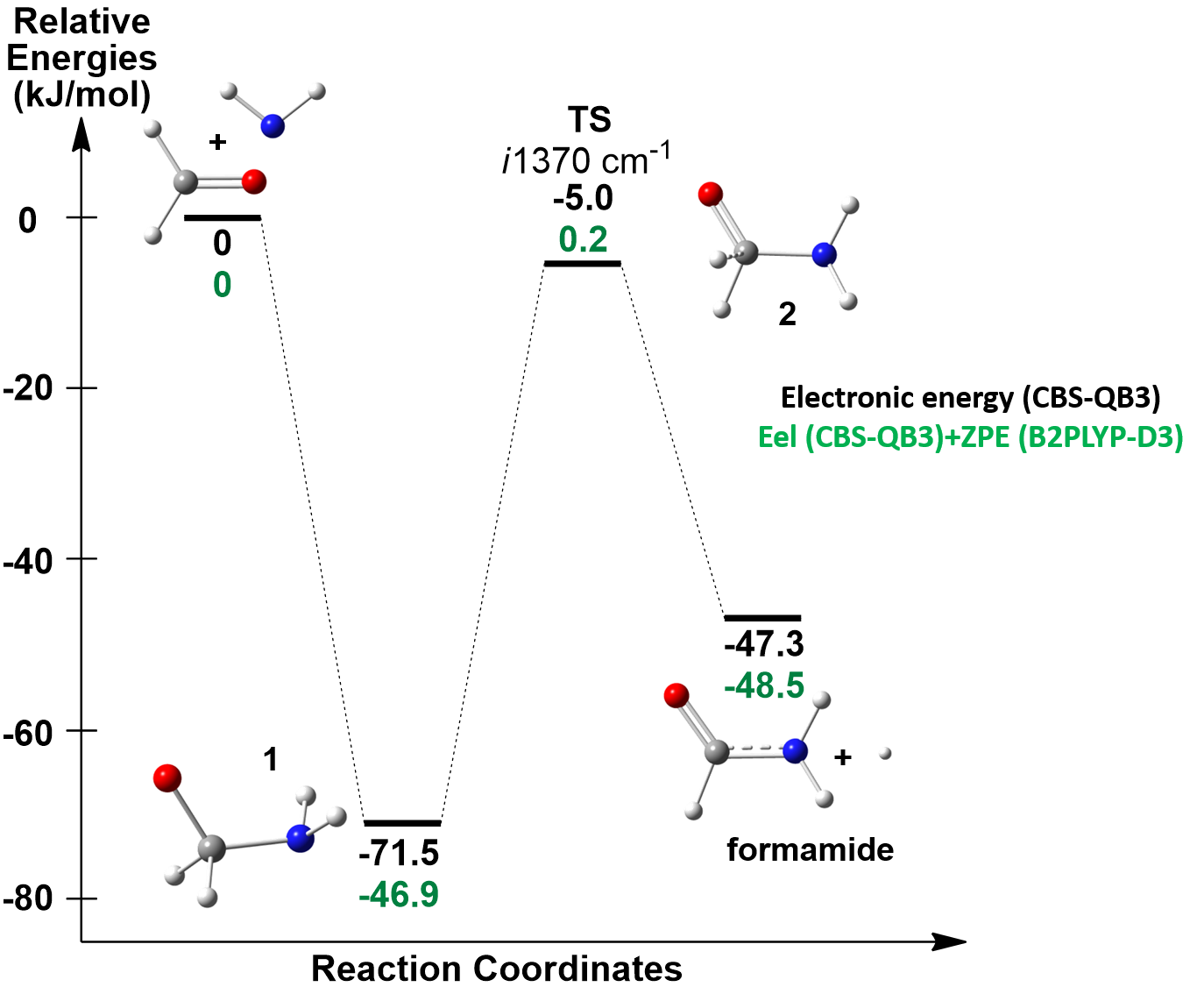}
        \caption{Proposed reaction path for formamide formation. The
          $y$ axis reports the relative electronic energies (black:
          CBS-QB3) and free energy differences at 0 K (green:
          B2PLYP-D3/m-aug-cc-pVTZ harmonic frequencies).}
    \label{fig:fig1}
\end{figure}

\begin{figure}
	\includegraphics[width=\columnwidth]{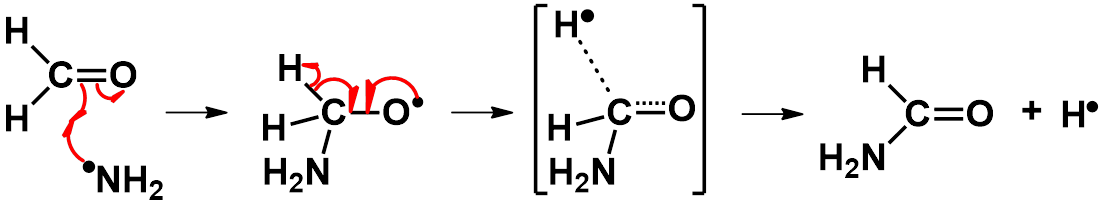}
    \caption{Possible radical mechanism for formamide formation.}
    \label{fig:fig2}
\end{figure}

Figure \ref{fig:fig1} depicts the possible path of the full CH$_2$O + NH$_2$ $\to$
NH$_2$CHO formation reaction and the relative electronic energies of
all the involved compounds, obtained at the CBS-QB3 level. Zero point
energies (ZPE) issuing from B2PLYP-D3BJ/m-aug-cc-pVTZ harmonic
frequencies were added to CBS-QB3 electronic energies to obtain the
$0K$ free energies (of course identical to enthalpies) also shown in
Figure \ref{fig:fig1}. The proposed pathway includes one intermediate (1) and one
transition state (2). In the first step of the reaction the NH$_2$
radical attacks the carbon atom of formaldehyde, leading to the
H$_2$C(NH$_2$)O radical 1, that is about $71$ kJ/mol more stable
than the precursors. This step is followed by hydrogen
loss from the carbon atom, leading to formamide together with H
radical, and ruled by the transition state 2, which exhibits a $67$
kJ/mol barrier (reduced to $46$ kJ/mol if ZPE is included). The
products were found to be $\sim 46$ kJ/mol more stable than the
reagents. The proposed radicalic mechanism is sketched in Figure
\ref{fig:fig2}. 

Alternative paths have been considered, including H migrations in
compound 1 (from C and N to O atoms), NH$_2$ attack on the oxygen atom
of formaldehyde or hydrogen abstraction from NH$_2$ which could have
led to NH$_3$ and HCO. However, all these reaction channels are ruled
by activation energies too high to be overcome in the interstellar
medium. A plausible tautomeric form of formamide was also
investigated, but once again the corresponding reaction channel is
closed under the typical conditions of the interstellar medium.

\subsection{Kinetics}

\begin{figure}
	\includegraphics[width=\columnwidth]{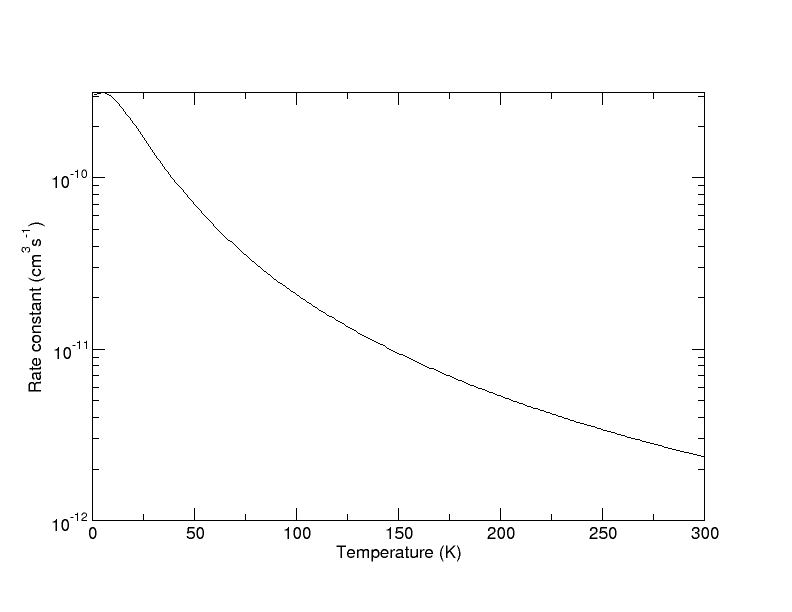}
        \caption{Total bimolecular rate constant for formamide
          formation as a function of temperature.}
    \label{fig:fig3}
\end{figure}

Figure \ref{fig:fig3} shows the variation of the computed rate constant for
formamide formation as a function of temperature up to 300 K. The
calculated rate constants have also been fitted, for temperatures from
10 K onwards, to an expression of the form:

\begin{equation}
  k(T)=A \times (T/300)^\beta e^{- \gamma / T}
\end{equation}
with $A = 2.6 \times 10^{-12}$ cm$^3$ s$^{-1}$, $\beta = -2.1$ and
$\gamma = 26.9$ K. The rate constant decreases monotonically as the
temperature increases.

Even though the activation energy governing H loss is marginally
higher than the energy of the reactants, this difference is so small
that, even at the lowest energies considered, tunneling leads to an
appreciable rate for hydrogen loss. In fact, at low energies (and
temperatures), H loss largely predominates over back-dissociation. The
rates of the two processes become equal around 10 K and then, as the
energy increases further, the rate of back-dissociation increases much
faster than the one for H loss due to the more rapid increase of the
density of states of the reactants (which include free rotations) and,
as a result, the rate for formamide formation drops.

\begin{figure}
	\includegraphics[width=\columnwidth]{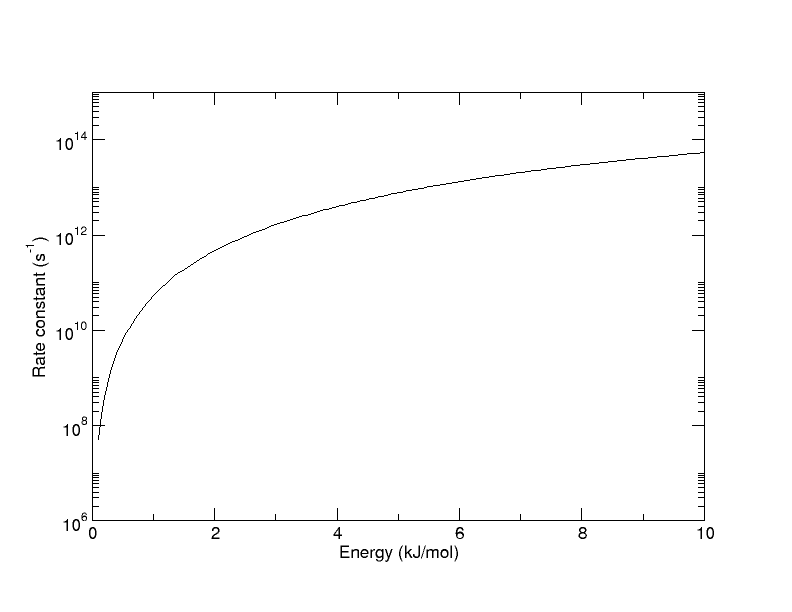}
        \caption{Unimolecular rate constant for dissociation to
          reactants of initial intermediate as a function of total
          energy.}
    \label{fig:fig4}
\end{figure}

Figure \ref{fig:fig4} shows the unimolecular rate constant and
back-dissociation as a function of energy.

\subsection{Astrochemical modelling}

In order to test the impact of the new rates on the predictions of the
formamide abundance, we run a series of models with the aim to
reproduce the observations in two different objects: the cold envelope
of the Sun-like protostar IRAS16293-2422 (Jaber et al. 2014) and the
molecular shock L1157-B2 (Mendoza et al. 2014). We chose these two
cases because they represent two extreme conditions, with the lowest
($\sim 6\times10^{-12}$) and highest ($\sim 2\times10^{-8}$) measured
abundance of formamide\footnote{Note that the abundances here are
  given with respect to the H atoms, while in the original articles
  they are with respect to H$_2$.}.
For the model, we used the time-dependent Nahoon
code\footnote{http://kida.obs.u-bordeaux1.fr.} (Wakelam et
al. 2012). It resolves the gas-phase chemical equation as a function
of the time, using the KIDA 2014 chemical network (Wakelam et
al. 2015), containing 489 species and 7499 reactions. We corrected the
rate coefficient of the NH$_2$ + H$_2$CO reaction according the
computations presented in the previous section.
In the following, we will discuss the two cases of IRAS16293-2422 and
L1157-B2 separately.

\begin{figure}
  \centering
  \includegraphics[width=0.9\columnwidth,angle=0]{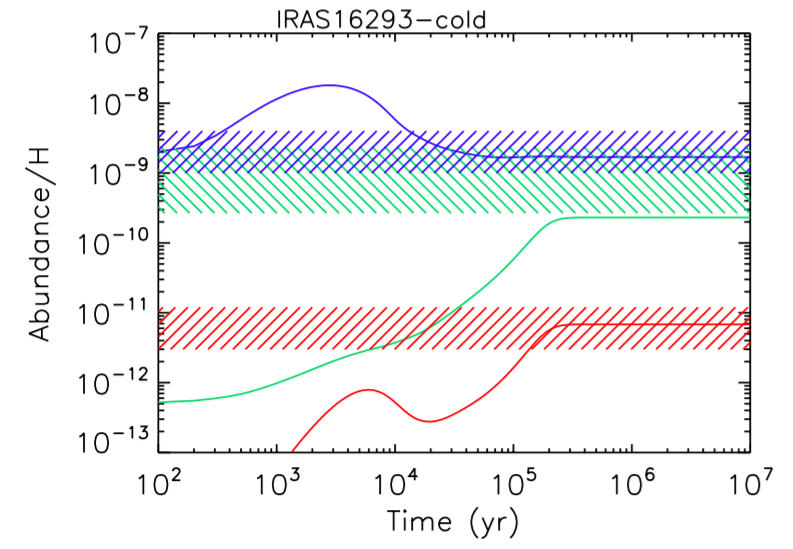}
 \caption{Abundances (with respect of H atoms) in the cold envelope of
 IRAS16293-2422 as a function of time (in years): H$_2$CO (blue),
 NH$_2$CHO (red) and NH$_2$ (green). The dashed boxes show the
 relevant measured values.}
  \label{I162932model-figure}
\end{figure}

\smallskip
\noindent {\it IRAS16293-2422, cold envelope:} IRAS16293-2422 is a
well studied Sun-like protostar of 22 L$_\odot$ (e.g. Crimier et al.
2010; Caux et al. 2011). Specifically, the density of the envelope
increases going inward with a power law of $\sim$1.5, from a H$_2$
density of $2 \times 10^5$ cm$^{-3}$ at 7000 AU. The temperature also
increases going inward from 10 K at the border to 100 K at 85 AU
(Crimier et al. 2010). Kahane et al. (2013) detected the formamide
using the unbiased spectral survey
TIMASSS\footnote{http://www-laog.obs.ujf-grenoble.fr/heberges/timasss/.}
(Caux et al. 2011). Successively, Jaber et al. (2014) modeled the
formamide line emission to disentangle the contribution from the cold
and the warm (hot corino) part of the IRAS16293-2422 envelope and
found that the average formamide abundance in the cold envelope is
$\sim 6\times10^{-12}$. For our chemical model, we used an average H
density of $2\times10^6$ cm$^{-3}$ and temperature of 20 K. We run the
model assuming that the elemental abundances in Wakelam et al. (2008;
Table 1, column EA2) are depleted by a factor 10 (C, O and N) and 100
(the heavier elements), in agreement with previous observations. With
these assumptions, the steady state abundance of NH$_2$CHO, H$_2$CO
and NH$_2$ result in excellent agreement with the observed values, as
shown in Fig. \ref{I162932model-figure}. Note that the implicit
assumption is that the cold envelope is formed by material previously
present in the placental molecular cloud, whose age is $\geq 10^6$ yr.

\begin{figure}
  \centering
  \includegraphics[width=0.9\columnwidth,angle=0]{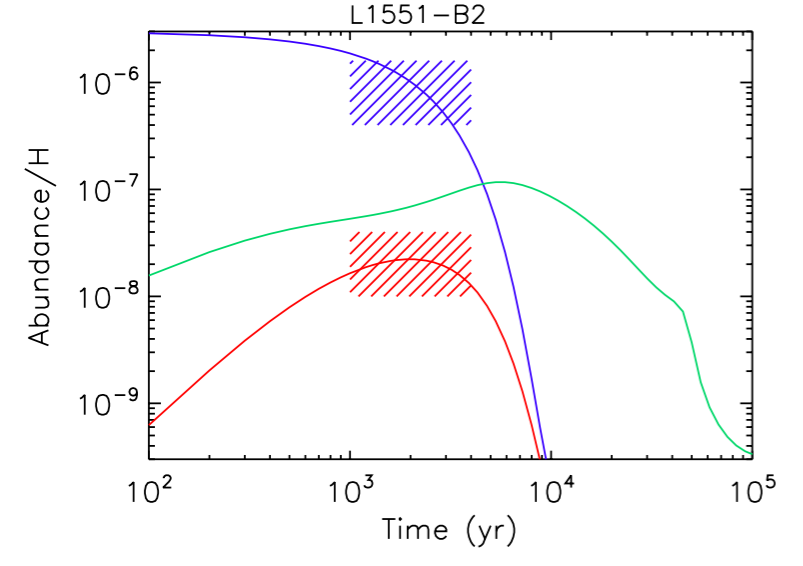}
 \caption{Abundances (with respect of H atoms) after the passage of
 the shock in L1157-B2 as a function of time (in years): H$_2$CO
 (blue), NH$_2$CHO (red) and NH$_2$ (green). The dashed boxes show the
 relevant measured values.}
  \label{L1551B2model-figure}
\end{figure}

\smallskip
\noindent {\it L1157-B2:} L1157-B2 is a molecular shock which is part
of the molecular outflow system emanating from L1157-mm (e.g. Gueth et
al. 1996; Lefloch et al. 2012). This system has extensively been
studied. Briefly, it is composed by two major ejection events, at the
origin of two molecular shock sites: B2, created by the first event,
and B1, created by the second one and which is spatially closer to the
central object. B1 is the best studied of the two molecular shocks, as
it has (also) been observed in the framework of two unbiased spectral
surveys programs, CHESS (Ceccarelli et al. 2010; {\it
  http://chess.obs.ujf-grenoble.fr/}) and ASAI (Lefloch, Bachiller et
al. in prep.; {\it http://www.oan.es/ASAI}), which provided the spectral
coverage of the 500-2000 GHz with Herschel and the 3, 2, and 1mm bands
with IRAM-30m, respectively. Mendoza et al.  (2014) measured the
formamide abundance in B2 and B1, equal to $\sim 2\times10^{-8}$ and
$\sim 1\times10^{-8}$, respectively.  Relevant to the simulations
reported here, Lefloch et al. (2012) provided constrains on the B1
density and temperature, $\sim 70$ K and $\sim 10^5$ cm$^{-3}$
respectively. Podio et al. (2014) used the line emission from several
molecular ions to further constrain the cosmic ray ionisation rate in
the region, $3\times 10^{-16}$ s$^{-1}$, a parameter important in
chemistry as it influences the timescale of the chemical evolution. We
adopted the above values also for the B2 molecular shock. The model,
as in previous works (Podio et al. 2014; Mendoza et al. 2014),
consists in a two-step procedure: (1) in the first step, the steady
chemical abundances are computed assuming that the elemental
abundances in Wakelam et al. (2008; Table 1, column EA2) are depleted
by a factor 10 (C, O and N) and 100 (the heavier elements); (2) in the
second step, CO, H$_2$O, H$_2$CO and NH$_3$\footnote{NH$_2$ is then
  synthesised by NH$_3$, which is the N- element iced major recevoir
  (e.g. Boogert et al. 2015).} are injected in the gas phase, to
simulate the passage of the shock, and the chemical evolution is
followed until a few $10^4$ yr, even though the age of the B1 shock is
evaluated to be around 2000 yr and the B2 a bit older. The CO and the
H$_2$O abundances (with respect to H) are $8\times 10^{-5}$ and
$1\times 10^{-4}$, respectively. We run several models varying the
H$_2$CO and NH$_3$ abundances.
Figure \ref{L1551B2model-figure} shows the abundance of NH$_2$CHO,
H$_2$CO and NH$_2$ obtained assuming H$_2$CO and NH$_3$ abundances
equal to $3\times 10^{-6}$ and $1\times 10^{-6}$, respectively, in
agreement with their measurements by Mendoza et al. (2014) and Tafalla
\& Bachiller (1995). The dashed areas show the measured values of
NH$_2$CHO and H$_2$CO\footnote{The upper limit on the NH$_2$ in
  Mendoza et al. (2014) strictly applies to B1, where the NH$_2$ were
  observed but not detected. In the case of B2, there are no specific
  observations.}.  The predictions obtained using the new values
perfectly reproduce the observations between 1000 and 3000 yr, the
estimated age of the shock. The predicted NH$_2$ abundance is about
5--9$\times 10^{-8}$.

We also run a few models for the specific case describing B1.  We can
reproduce the measured formamide abundance and predict a NH$_2$
abundance of $\sim 10^{-8}$. From the Mendoza et al. (2014) upper
limit, we derive a 3 $\sigma$ level upper limit on the NH$_2$
abundance of $\sim 3\times 10^{-9}$, which is a factor $\sim 3$ lower
than the predicted one. We caution that these upper limits suffer of
the uncertainty on the excitation temperature, and, therefore, further
dedicated observations should be carried out to better constraint the
predictions.

In summary, the new rate for the reaction H$_2$CO and NH$_2$
$\rightarrow$ NH$_2$CHO allows gas phase reactions fully justify the
observed formaldehyde abundances, both in the cold and warm gas
sources, without the need to invoke specific grain-surface reactions
for that.

\section{Conclusion and perspectives}

In this paper, we have provided new insights concerning the formation 
of formamide in the interstellar medium. Our computations allowed us 
to propose a reaction path combined to a plausible mechanism concerning 
this formation. Indeed, the first addition step does not involve any 
barrier and can therefore occur in space. RRKM calculations confirmed 
the feasibility of this reaction since once this first addition is done, 
formamide formation largely predominates over back-dissociation at low 
energies. Moreover, we tested the impact of these new 
kinetic data on the prediction of formamide abundance, comparing it 
with observations in the cold envelope of the Sun-like protostar 
IRAS16293-2422 and the molecular shock L1157-B2 (lowest and highest 
measured abundance of formamide). The results obtained from these 
simulations are in excellent agreement with the observation and, therefore, 
confirmed our computational protocol.

A more general conclusion is that neutral-neutral gas-phase reactions
can account for the formation of relatively complex organic molecules
even under the extreme conditions of ISM. Grain-surface reactions are
often called into play to explain the formation of complex organic
molecules because of supposedly missing formation routes in the gas
phase. However, not all the possible gas-phase routes have been
actually explored, either in laboratory experiments or theoretically,
or correctly included in the astrochemical networks and models (see,
for instance, Balucani et al. 2015). Other studies of critical and yet
unexplored neutral neutral gas phase reactions performed with the same
theoretical methods employed here can help to fulfill the gap of the
missing reactions leading to complex organic molecules, especially
when those reactions cannot be easily investigated in laboratory
experiments.

\section*{Acknowledgements}
We wish to thank J-C. Loison for fruitful discussions.  We acknowledge
the financial support from the COST Action CM1401 ``Our Astrochemical
History''. NB acknowledges the financial support from the Universit\'e
Grenoble Alpes and the Observatoire de Grenoble. The research leading
to these results has received funding from the European Research
Council under the European Union's Seventh Framework Programme
(FP/2007-2013) / ERC Grant Agreement n. [320951].

%%%%%%%%%%%%%%%%%%%%%%%%%%%%%%%%%%%%%%%%%%%%%%%%%%


% Don't change these lines
\bsp	% typesetting comment

\label{lastpage}


\begin{thebibliography}{}

\bibitem[\protect\citeauthoryear{Adande et al.}{2011}]{Adande2011}  Adande G. R., Woolf N. J. G., Ziurys L. M., 2011, Astrobiology, 13, 439

\bibitem[\protect\citeauthoryear{Balucani et al.}{2015}]{Balu2015}
  Balucani N., Ceccarelli C., Taquet V., 2015, MNRAS 449, L16

\bibitem[\protect\citeauthoryear{Bisschop et al.}{2007}]{Bisschop2007} Bisschop S. E. et al., 2007, A\&A, 465, 913

\bibitem[\protect\citeauthoryear{Biver et al.}{2014}]{Biver2014} Biver et al., 2014, A\&A, 566, L5

\bibitem[\protect\citeauthoryear{Bockelee-Morvan et al.}{2000}]{Bockelee-Morvan2000} Bockelee-Morvan D. et al., 2000, A\&A, 53, 1101

\bibitem[\protect\citeauthoryear{Caselli \& Ceccarelli}{2012}]{Caselli2012} Caselli P. \& Ceccarelli C., 2012, A\&A Rev, 20, 56

\bibitem[\protect\citeauthoryear{Dunning}{1989}]{Dunning1989} Dunning T. H. 1989, J. Chem. Phys. 90, 1007

\bibitem[\protect\citeauthoryear{Frisch et al.}{2013}]{Frisch2013}
  Frisch, M. J. et al. Gaussian09 GDVH32, 2013, GDVH32

\bibitem[\protect\citeauthoryear{Garrod et al.}{2008}]{Garrod2008} Garrod R. T., Weaver S. L. W., Herbst E., 2008, ApJ, 682, 283

\bibitem[\protect\citeauthoryear{Garrod}{2013}]{Garrod2013} Garrod R. T., 2013, ApJ, 765, 60

\bibitem[\protect\citeauthoryear{Goerigk \& Grimme}{2011}]{GoerigkGrimme2011} Goerigk L., Grimme S. 2011, J. Chem. Theory Comput. 7, 291-309

\bibitem[\protect\citeauthoryear{Grimme}{2006}]{Grimme2006} Grimme S. J. 2006 Chem. Phys. 124, 034108

\bibitem[\protect\citeauthoryear{Grimme et al.}{2011}]{Grimme2011}
  Grimme S., Ehrlich S., Goerigk L., 2011, J. Comput. Chem. 32, 1456-1465

\bibitem[\protect\citeauthoryear{Gueth et al.}{1996}]{Gueth1996} Gueth
  F. et al., 1996, A\&A 307, 891

\bibitem[\protect\citeauthoryear{Hollis et al.}{2006}]{Hollis2006} Hollis J. M., et al. 2006, ApJ, 643, L25

\bibitem[\protect\citeauthoryear{Jones et al.}{2011}]{Jones2011} Jones B. M., Bennett C. J., \& Kaiser R. I. 2011, ApJ, 734, 78

\bibitem[\protect\citeauthoryear{Kahane et al.}{2013}]{Kahane2013}Kahane C., Ceccarelli C., Faure A., Caux E., 2013, ApJ, 763, L38

\bibitem[\protect\citeauthoryear{Lefloch et al.}{2012}]{Lefloch2012}
  Lefloch B. et al., 2012, ApJL, 757, L25

\bibitem[\protect\citeauthoryear{Leonori et al.}{2009}]{Leonori2009}
  Leonori F., et al. 2009, J. Phys. Chem. A  113, 15328-15345

\bibitem[\protect\citeauthoryear{Leonori et al.}{2013}]{Leonori2013}
  Leonori F. et al., 2009, J. Chem. Phys. 2013, 138, 024311

\bibitem[\protect\citeauthoryear{Li \& Lu}{2002}]{Li2002}Li Q. S., \& Lu R. H., 2002, JPCA, 106, 9446

\bibitem[\protect\citeauthoryear{Lopez-Sepulcre et al.}{2015}]{Lopez-Sepulcre2015}Lopez-Sepulcre A. et al., 2015, MNRAS 449, 2438

\bibitem[\protect\citeauthoryear{Mendoza et al.}{2014}]{Mendoza2014}Mendoza E. et al., 2014, MNRAS, 445, 151

\bibitem[\protect\citeauthoryear{Montgomery et al.}{2000}]{Montgomery2000} Montgomery J. A., Frisch M. J., Ochterski J. W., Petersson G. A. 2000, J. Chem. Phys. 112, 6532

\bibitem[\protect\citeauthoryear{Noble et al.}{2015}]{Noble2015}Noble J. A. et al., 2015, A\&A, 576, 91

\bibitem[\protect\citeauthoryear{Ochterski et al.}{1996}]{Ochterski1996} Ochterski J. W., Petersson G. A., Montgomery J. A., 1996, J. Chem. Phys. 104, 2598

\bibitem[\protect\citeauthoryear{Papajak et al.}{2009}]{Papajak2009} Papajak, E. et al. 2009, J. Chem. Theory Comput. 5, 1197-1202

\bibitem[\protect\citeauthoryear{Quan et al.}{2007}]{Quan2007} Quan D., Herbst E. 2007, A\&A, 474, 521

\bibitem[\protect\citeauthoryear{Redondo et al.}{2014a}]{Redondo2014a}
  Redondo et al., 2014a, ApJ 793, 32

\bibitem[\protect\citeauthoryear{Redondo et al.}{2014b}]{Redondo2014b}
  Redondo et al., 2014b, ApJ, 780, 181

\bibitem[\protect\citeauthoryear{Rubin et al.}{1971}]{Rubin1971} Rubin
  R. H. et al. 1971, ApJ, 169, L39

\bibitem[\protect\citeauthoryear{tafalla}{1995}]{tafa95} Tafalla M.,
  Bachiller R., 1995, ApJL 443, L37

\bibitem[\protect\citeauthoryear{Turner}{1989}]{Turner1989} Turner B. E. 1989, ApJS, 70, 539

\bibitem[\protect\citeauthoryear{Vazart et al.}{2015}]{Vazart2015}
  Vazart F.et al., 2015, ApJ in press

\bibitem[\protect\citeauthoryear{Yamaguchi et
    al.}{2012}]{Yamaguchi2012} Yamaguchi T. et al., 2012, PASJ, 64, 105


\end{thebibliography}
\end{document}